\documentclass[twoside,preprintnumbers,amsmath,amssymb,pacs,shownopacs,nofootinbib,notitlepage]{revtex4-2}

\usepackage{graphicx,amsmath,amsfonts,amssymb,dcolumn,amsthm,euscript,braket}
\usepackage{slashed}
\usepackage{xcolor}
\usepackage[symbol*]{footmisc}
\usepackage{caption}
\usepackage{subcaption}
\usepackage{subcaption}
\usepackage[margin=10pt,font=small,labelfont=bf,
labelsep=endash, format=hang, justification=raggedright]{caption}
\usepackage{float}
\usepackage[makeroom]{cancel}
\usepackage[export]{adjustbox}
\usepackage{tikz}       % Powerful drawing language
\usetikzlibrary{decorations.markings}
\usepackage{tikz-feynman}

\begin{document}
\tikzset{->-/.style={decoration={
  markings,
  mark=at position #1 with {\arrow[>=latex]{>}}},postaction={decorate}}}
 
\tikzset{-<-/.style={decoration={
  markings,
  mark=at position #1 with {\arrow[>=latex]{<}}},postaction={decorate}}}

\title{{\bf Landau-Khalatnikov-Fradkin transformations for the two loop massless quark propagator}}

\author{P. Dall'Olio$^{\dag}$, T. De Meerleer$^{\ddag}$, D.~Dudal$^{\ddag, \sharp}$, S.~P.~Sorella$^\ast$, A.~Bashir$^\S$}
\email{pietro@matmor.unam.mx, timdemeerleer07@gmail.com, david.dudal@kuleuven.be, silvio.sorella@gmail.com, adnan.bashir@umich.mx}
\affiliation{$\dag$Centro de Ciencias Matem\'{a}ticas, Unam - Campus Morelia, Antigua Carretera a P\'atzcuaro 8701
Col. Ex Hacienda San Jos\'e de la Huerta
Morelia, Michoac\'an  58089, M\'exico  \\
$\ddag$ KU Leuven Campus Kortrijk--Kulak, Department of Physics, Etienne Sabbelaan 53 bus 7657, 8500 Kortrijk, Belgium \\
$\sharp$ Ghent University, Department of Physics and Astronomy, Krijgslaan 281-S9, 9000 Gent, Belgium
\\$^\ast$ Instituto de F\'isica Te\'orica, Rua S\~ao Francisco Xavier 524, 20550-013, Maracan\~a, Rio de Janeiro, Brasil \\  $\S$ Instituto de F\'{i}sica y Matem\'{a}ticas, Universidad Michoacana de San Nicol\'{a}s de Hidalgo,
Edificio C-3, Ciudad Universitaria, Morelia, Michoac\'{a}n 58040, M\'{e}xico}

\begin{abstract}
Landau-Khalatnikov-Fradkin transformations (LKFTs) yield the gauge dependence of correlation functions within the class of linear covariant gauges. We derive the LKFT for the quark propagator and explicitly evaluate it up to the two loop level in the chiral limit. Although the number of diagrams to be evaluated is significantly larger than with the conventional computational scheme, the diagrams are simpler in nature,
thereby leading to a considerably faster evaluation of the gauge dependent part than naively expected. Finally, we
also resum the LKFT generated terms and compare our results with earlier work in the literature.
\end{abstract}

\maketitle

\section{Introduction}
The Landau-Khalatnikov-Fradkin transformations (LKFTs) \cite{Landau:1955,Fradkin:1955} are a set of identities that relate the expressions of a correlation function evaluated in different linear covariant gauges. In particular, they allow to split the content of a correlation function into its gauge independent and its gauge dependent part, this last one parametrized by the gauge parameter $\alpha$. 

In a set of previous works \cite{DeMeerleer:2018txc,DeMeerleer:2020} we performed a detailed study of the LKFT for the two-point correlation function $\langle A^a_\mu(p) A^b_\nu(-p)\rangle_{\alpha}$ of the non-Abelian gauge field $A^a_\mu$. The main tool of our approach was the use of a composite transverse gauge invariant quantity $A^{ah}_\mu$ built up by means of an auxiliary (scalar) Stueckelberg field $\xi^a$. More precisely, we define
\begin{equation} \label{Ahloc}
A_\mu^h = h^\dagger A_\mu h+\frac{i}{g}h^\dagger \partial_\mu h, \qquad h=e^{igT^a \xi^a}.
\end{equation}
Its gauge invariance is guaranteed by the action of the SU($N$) gauge transformation $U=e^{igT^a \alpha^a}$ on $h$
\begin{equation} \label{Uinv}
\begin{split}
& h^U\equiv U^\dagger h\,, \qquad A^U_\mu = U^\dagger A_\mu U + \frac{i}{g} U^\dagger \partial_\mu U \\
&\Longrightarrow (A^h_\mu)^U = (h^\dagger)^U A_{\mu}^U h^U + \frac{i}{g}(h^\dagger)^U \partial_\mu h^U = A^h_\mu,
\end{split}
\end{equation}
where the gauge transformed fields are denoted with the superscript $U$. The transversality of $A_\mu^h$, $\partial_\mu A_\mu^h=0$, which will be imposed by means of a multiplier, is crucial to obtain a renormalizable composite field operator \cite{Capri:2016a2, Capri:2017ren, Capri:2018ir}. 

 The employment of the quantity $A^{ah}_\mu$ enabled us \cite{DeMeerleer:2018txc} to establish the LKFT relating the correlation function $\langle A^a_\mu(p) A^b_\nu(-p)\rangle_{\alpha}$ computed for a generic value of the gauge parameter $\alpha$ to the corresponding two-point function in the Landau gauge, i.e.~$\langle A^a_\mu(p) A^b_\nu(-p)\rangle_{\alpha=0}$. Besides, we were able to connect the LKFTs to the so called Nielsen identities \cite{Nielsen:1975fs,Piguet:1984js,Quadri:2014jha} which follow from the generalized Slavnov-Taylor identities, see \cite{DeMeerleer:2020}.

The present work aims at facing the issue of the LKFT for the quark propagator. More precisely, by making use of the gauge invariant composite fermionic field $\psi^h$, see eqs.~\eqref{psih},\eqref{psih_expansion}, we obtain the relationship between the quark propagator $\langle {\bar \psi}^{i}(p) \psi^{j}(-p) \rangle_{\alpha}$ evaluated for a generic $\alpha$ and the same correlator in the Landau gauge, namely $\langle {\bar \psi}^{i}(p) \psi^{j}(-p) \rangle_{\alpha=0}$. We carry out the calculation till two loop order by making use of dimensional regularization in the $\overline{\text{MS}}$ scheme.

Although we focus on the LKFT for the quark propagator, it is worth mentioning that the setup outlined in  \cite{DeMeerleer:2018txc,DeMeerleer:2020} can be applied to obtain the non-Abelian LKFT for an arbitrary $n$-point correlation function built up with the
fields $(A^a_\mu, {\bar \psi}^{i}, \psi^{i})$. We stress that, relating the correlation function in an arbitrary gauge to its counterpart evaluated in Landau gauge, is in fact equivalent to connecting the expressions for the correlation function corresponding to different gauge parameters.

Up to recently, the studies concerning the LKFTs were confined to the Abelian case of QED, and in particular to the electron propagator, using the standard tools of quantum field theory as well as the quantum mechanical approach of the worldline formalism \cite{Bashir:2002sp,Ahmad:2016dsb,Ahmadiniaz:2015kfq,Ahmadiniaz:2020msv,Jia:2017pl,Jia:2017pr,Kotikov:2019,Gusynin:2020cra}.  In fact,
two-loop electron propagator has also been discussed in different dimensions invoking lkft in qed 
\cite{Bashir:2000ur,Bashir:2000rv,Bashir:1999bd,Bashir:2004hh}. 
 An exception being \cite{Aslam:2016} where it was attempted to derive the LKFT for the quark propagator. In that work, though, the auxiliary field was heuristically considered free like in the Abelian case, which is not the case as shown from first principles in \cite{DeMeerleer:2018txc} and in the following. It is in fact the non-vanishing of the interactions among the auxiliary field and the other fields that makes the evaluation of the LKFTs in the non-Abelian case quite involved.  

The paper is organized as follows. In Section II we provide a brief overview of the framework put forward in \cite{DeMeerleer:2018txc,DeMeerleer:2020}. In Section III the LKFT for the quark propagator are derived till two loops, as summarized in eq.~\eqref{lkft}, eq.~\eqref{lkf1l} and eq.~\eqref{result}. As a consistency check of the whole procedure, the gauge dependence of the quark renormalization constant  $Z_2$ is correctly re-obtained to this order of approximation.

\section{Classical action}
We consider the following classical action in Euclidean space \cite{Capri:2016a2, Capri:2017ren, DeMeerleer:2018txc}
\begin{equation} \label{action}
S= S_{\mathrm{QCD}} + S_{\mathrm{gf}} + S_\mathrm{h},
\end{equation}
where $S_{\mathrm{QCD}}$ is the usual gauge invariant QCD action which encodes the dynamics of quarks and gluons
\begin{equation}
S_{\mathrm{QCD}} = \int d^D x \left[\frac{1}{4} F^a_{\mu \nu}F^a_{\mu \nu} + \bar \psi_f \left(\gamma_\mu D_\mu\right)\psi_f \right],
\end{equation}
where $N_f$ flavours of massless quarks are considered. $F^a_{\mu \nu} = \partial_\mu A^a_\nu - \partial_\nu A^a_\mu + g f^{abc}A^b_\mu A^c_\nu$ is the gluon field strength tensor and $D_\mu = \partial_\mu - i g T^a A^a_\mu$ is the covariant derivative, $T^a$ being the generators of the SU($N$) group which satisfy the Lie algebra $[T^a, T^b]=if^{abc}T^c$, with normalization $\text{Tr}(T^a T^b)=\frac{\delta^{ab}}{2}$.  $S_\mathrm{gf}$ includes the gauge-fixing terms in linear covariant gauges
\begin{equation} \label{eq:gf}
S_{\mathrm{gf}} = \int d^D x \left[ \frac{\alpha}{2}b^a b^a + i b^a \partial_\mu A^a_\mu + \bar c^a \partial_\mu D^{ab}_\mu c^b\right],
\end{equation}
where $b^a$ is the Nakanishi-Lautrup auxiliary field which implements the gauge condition  \cite{Nakanishi:1966}, $\alpha$ is the gauge parameter, $c^a(x)$ and $\bar c^a(x)$ are the anti-commuting ghost fields which yield the exponential representation of the Jacobian arising in the gauge fixing procedure, and $D^{ab}_\mu=\partial_\mu \delta^{ab}-gf^{abc}A^c_\mu$ is the covariant derivative in the adjoint representation. The \emph{Landau} gauge corresponds to $\alpha = 0$, for which the auxiliary field $b^a$ strictly enforces the transversality condition $\partial_\mu A^a_\mu=0$.

This standard gauge-fixed QCD action is augmented by the term
\begin{equation} \label{S_h}
S_\mathrm{h} = \int d^D x \left[\tau^a \partial_\mu A^{h,a}_\mu +\bar \eta^a\, \partial_\mu D^{ab}_\mu(A^h) \eta^b\right],
\end{equation}
where $A^h_\mu$ is forced to be transverse through the introduction of the auxiliary field $\tau^a$ in $S_h$. It is the localized representation of the gauge invariant non-local operator that minimizes the functional $\int d^D x\, \mathrm{Tr}\, A^U_\mu A^U_\mu$ along a gauge orbit, given by (see for instance \cite{Capri:2015brs})
\begin{equation} \label{Ahnonloc}
A^h_\mu = \left(\delta_{\mu \nu} - \frac{\partial_\mu \partial_\nu}{\partial^2}\right) \left(A_\nu -ig\left[\frac{1}{\partial^2}\partial_\sigma A_\sigma, A_\nu \right] +\frac{ig}{2} \left[\frac{1}{\partial^2}\partial_\sigma A_\sigma, \partial_\nu \frac{1}{\partial^2}\partial_\sigma A_\sigma  \right] + {\cal O}(A^3)\right).
\end{equation}
Each term in this non-local expression contains at least one factor of the gauge field divergence, which makes it explicit how in Landau gauge ($\partial_\mu A_\mu=0$) $A^h_\mu = A_\mu$.
By expanding \eqref{Ahloc} in powers of $\xi$ and imposing the transversality condition, one can iteratively solve for the Stueckelberg field  $\xi$,
\begin{equation} \label{xieom}
\xi= \frac{1}{\partial^2}\partial_\mu A_\mu+ i\frac{g}{\partial^2}\left[\partial_\mu A_\mu,\frac{\partial_\sigma}{\partial^2}A_\sigma\right]+i\frac{g}{\partial^2}\left[A_\mu,\frac{\partial_\mu}{\partial^2}\partial_\sigma A_\sigma\right]+i\frac{g}{2\partial^2}\left[\frac{\partial_\mu}{\partial^2}A_\mu,\partial_\sigma A_\sigma\right]+ {\cal O}(A^3).
\end{equation}
and recover the non-local representation given in \eqref{Ahnonloc}. The introduction of a new pair of Grassmannian ghost fields $\eta$ and $\bar \eta$ in \eqref{S_h} is required in order to take care of the non-trivial Jacobian arising through the dependence of $A^h_\mu$ on the Stuckelberg field. In fact, if one integrates back the auxiliary field $\tau$ and the pair of ghost fields, one is left with the path integration over the Stueckelberg field, yielding, up to an irrelevant global factor,
\begin{equation}
\int D\xi \, \delta \! \left(\partial_\mu A^h_\mu\right) \det \left(-\partial_\mu D_\mu( A^h_\mu )\right) = 1.
\end{equation}
We point out that this result is based on the perturbative solution of the $\tau$-equation of motion, which results in the \emph{unique perturbative} series solution \eqref{xieom}. Beyond perturbation theory, this uniqueness might not prevail. This is the same identity introduced in the Faddeev-Popov gauge-fixing procedure. This means that the insertion of $S_h$ does not change the physical content of the theory. It leads to a classical action that is non-polynomial, given the infinite number of terms obtained by expanding \eqref{Ahloc} in terms of $\xi$,
\begin{equation} \label{Ahexp}
A^{h,a}_\mu = A^a_\mu - D^{ab}_\mu \xi^b -\frac{g}{2} f^{abc} \xi^b D^{cd}_\mu \xi^d -\frac{g^2}{3!} f^{abd}f^{dce} \xi^b \xi^c D^{ef}_\mu \xi^f + {\cal O}(\xi^4).
\end{equation}
Despite this fact, we emphasize that the Lagrangian is still local since each term in \eqref{Ahexp} is comprised of at most one derivative of $\xi$. This feature  is easily understood through dimensional analysis, the Stueckelberg field being dimensionless. This scenario shares similarities with the one encountered in the non-linear sigma models. For completeness, one can check that the explicit solution \eqref{Ahnonloc} is obtained from substituting the formal series solution \eqref{xieom} into the expression \eqref{Ahexp}.

Due to the gauge invariance of $A^h_\mu$, the new term represented by $S_h$ clearly does not spoil the standard BRST symmetry of the classical action; while the new fields $\tau$, $\eta$ and $\bar \eta$ are BRST singlets, the BRST transformation of the Stueckelberg field can be obtained iteratively from the transformation of $h$, ($sh^{ij}=-ig c^a (T^a)^{ik} h^{kj}$), yielding
\begin{equation}
s \xi^a = -c^a + \frac{g}{2}f^{abc}c^b\xi^c-\frac{g^2}{12}f^{abc}f^{bde}c^d\xi^e\xi^c + {\cal O}(\xi^3).
\end{equation}
The operator $A^h_\mu$ corresponds to a dressed gauge invariant gluon operator, \cite{Lavelle:1995ty}.

The introduction of the gauge invariant operator $A^h_\mu$ makes it straightforward to accommodate a dimension $d=2$ gluon operator in a gauge invariant fashion. It is achieved by adding the term $\frac{1}{2}m^2 A^{h,a}_\mu A^{h,a}_\mu$ to the action. This term takes into account the non-perturbative formation of a dimension two gluon condensate \cite{Chetyrkin:1999, Gubarev:2001as, Gubarev:2001, Boucaud:2001, Verschelde:2001} which is responsible for the infrared saturation of the gluon propagator, a behavior emerged in different continuum approaches \cite{Cornwall:1982, Aguilar:2004,Aguilar:2008xm, Dudal:2008,Ayala:2012pb,Siringo:2015wtx} and observed in lattice simulations \cite{Cucchieri:2007, Cucchieri:2008, Bogolubsky:2007, Oliveira:2012eh}.
This mass incorporates non-trivial infrared corrections to correlation functions, but it does not imply the presence of physical mass poles in the gauge boson sector. Indeed, the corresponding propagator still exhibits positivity violation \cite{Reinosa:2017qtf}, in accordance with lattice data \cite{Silva:2013rrr}.  The unitary question still needs to be addressed in terms of the gauge invariant bound state spectrum (glueballs, mesons and hybrids), about which we have nothing to say here.

It is noteworthy that, whether the $A_\mu^hA_\mu^h$ mass operator is included or not, action \eqref{action} defines a quantum theory which is renormalizable to all orders in perturbation theory \cite{Capri:2016a2, Capri:2017ren, Capri:2018ir}. This is quite remarkable given the infinite number of interactions present. For this fundamental property to hold, the transversality of the field $A^h_\mu$ is the key factor which distinguishes this formulation from the original non-power counting renormalizable Stueckelberg action \cite{Ruegg:2004}.

In particular, in Landau gauge, where $A^h_\mu= A_\mu$, this gauge invariant massive term will reduce to the massive extension of Yang-Mills theory, known as Curci-Ferrari model, which has recently been proved very successful in reproducing lattice results for the correlation functions \cite{Tissier:2011,Gracey:2019xom,Pelaez:2021tpq}.
Since we are merely interested in reproducing the correct gauge covariance at a perturbative level for now, we will not dwell on the effects of this massive operator.

However, the study of the LKFT for the quark propagator beyond perturbation theory might have several applications as, for example, the characterization of the $\alpha$-dependence of the so-called quark mass function, a quantity which is under current
intensive investigations from both theoretical approaches: e.g.~Schwinger-Dyson equations \cite{Alkofer:2008tt,Aguilar:2010cn,Bashir:2012fs}, Functional Renormalization Group \cite{Braun:2014ata,Cyrol:2017ewj}, effective massive models \cite{Barrios:2021cks}, as well as from numerical lattice computations \cite{Oliveira:2018lln}. We will come back to this issue in future work, where we will generalize the content of the current paper to that of the massive gluon/quark case.

\section{LKFT for the quark propagator}

In order to derive the covariant transformation of the quark propagator, we are going to use the gauge invariance of the composite fermion operator
\begin{equation} \label{psih}
\begin{split}
&\psi^h = h^\dagger \psi, \qquad \bar \psi^h = \bar \psi h, \\
\Longrightarrow &(\psi^h)^U = (h^\dagger)^U \psi^U=h^\dagger U U^\dagger \psi = \psi^h, \quad (\bar \psi^h)^U = \bar \psi^h,
\end{split}
\end{equation}
which can be expanded in terms of the Stuckelberg field $\xi$ as
\begin{equation}
\begin{split}
(\psi^h)^{i} &=  \psi^i -ig \xi^{a}(T^a)^{ij} \psi^j - \frac{g^2}{2}\xi^a \xi^b (T^a)^{ik}(T^b)^{kj} \psi^j + {\cal O}(\xi^3) \\
& = \psi^i + \Delta_1^i +  \Delta_2 ^i + {\cal O}(\xi^3) \label{psih_expansion},
\end{split}
\end{equation}
where in the last line we introduced a compact notation for the composite operators in the expansion, with the subindex in the $\Delta$ terms indicating the number of Stueckelberg fields present in the corresponding term, which coincides with the power of the coupling constant attached. An analogous expansion occurs for $\bar \psi^h$. As for the gluon, this procedure actually corresponds to dressing the fermion operator, making it explicitly gauge invariant, see again \cite{Lavelle:1995ty}.

Since the correlation function of a product of gauge invariant (and therefore BRST invariant) fields does not depend on the gauge parameter $\alpha$ \cite{Capri:2016a2, Capri:2017ren, Capri:2018un}, we have
\begin{equation} \label{gaugeind}
\langle  \bar \psi^{h} (x) \psi^{h}(y)  \rangle_{\alpha}  =
\langle \bar \psi^{h} (x) \psi^{h}(y) \rangle_{\alpha = 0} = \langle \bar \psi (x) \psi(y) \rangle_{\alpha = 0},
\end{equation}
where in the last identity we used the fact that in Landau gauge ($\alpha = 0$) the correlation function of the composite operators reduces to the correlation function of the usual elementary fields, since in this particular gauge the Stueckelberg field $\xi$ is forced to be zero on-shell \cite{Capri:2018un}, implying $A^h_\mu = A_\mu$ and $\psi^h (\bar \psi^h) = \psi (\bar \psi)$. This can also be observed by looking at the form of the Stueckelberg propagator (see below, $\langle \xi(p) \xi(-p)  \rangle= \frac{\alpha}{p^4}$), which vanishes in the Landau gauge.

Expanding the lhs of \eqref{gaugeind} in terms of the quark fields and the Stueckelberg field, one obtains
\begin{equation} \label{lkft}
 \langle \bar \psi^i (x) \psi^j(y) \rangle_{\alpha} =  \langle \bar \psi^i (x) \psi^j(y) \rangle_{\alpha = 0} + {\cal R}^{ij}_{\alpha}(x, y),
\end{equation}
where ${\cal R}^{ij}_{\alpha}(x,y)$, the gauge dependent part of the propagator, is given by the following terms, up to $\xi^2$ (we will omit the $\alpha$ subindex in the expectation values from now on)
\begin{align}
\label{rest}
%{\cal R}^{ij}_{\alpha}(x, y) &= ig T^{a}_{jk}\langle \bar \psi^i(x) \psi^k(y) \xi^a(y) \rangle_\alpha - igT^{a}_{ki}\langle \bar \psi^k(x) \psi^j(y) \xi^a(x) \rangle_\alpha
{\cal R}^{ij}_{\alpha}(x, y) &= - 2\langle \bar \psi^i(x)  \Delta_1^j(y) \rangle - 2\langle \bar \psi^i(x) \Delta_2^j(y) \rangle - 2\langle \bar \Delta_1^i(x) \Delta_2^j(y) \rangle -\langle \bar \Delta^i_1(x) \Delta^j_1(y) \rangle-\langle \bar \Delta^i_2(x) \Delta^j_2(y) \rangle,
\end{align}
where we used the symmetry $\langle \bar \psi^i(x)  \Delta_{1,2}^j(y) \rangle = \langle \bar \Delta_{1,2}^i(x)  \psi^j(y)  \rangle$ and $\langle \bar \Delta^i_1(x) \Delta^j_2(y) \rangle = \langle \bar \Delta^i_2(x) \Delta^j_1(y) \rangle$ inherited from the full fermion propagator.

In order to check the quark LKFT at the two loop level one might think that in the expansion \eqref{psih_expansion} the terms up to $O(g^4)$ should be kept. However, the third and fourth order terms would give rise only to Feynman diagrams containing tadpoles of the massless Stueckelberg field, which vanish when the loop integrals are made finite using dimensional regularization. This would not be the case if a mass scale were introduced as in \cite{Capri:2018ir}, in order to regularize the possible infrared divergences, since it transforms the Stueckelberg propagator into a massive propagator, making the tadpoles diagrams
non-vanishing. However, as it turns out, there are no such infrared divergences in dimensional regularization for the diagrams considered so far.

The composite operators inside the correlation functions in \eqref{rest} are inserted in the diagrams by adding to the classical action external sources attached to the composite operators
\begin{equation}
\label{source}
S_J = \int d^D x \left( \bar J_q^1(x)\Delta_1(x) + \bar J_q^2(x)\Delta_2(x) + \bar \Delta_1(x) J_q^1(x) + \bar \Delta_2(x) J_q^2(x) \right) \;,
\end{equation}
and then evaluating the two-point correlation functions between the fermionic fields and the external sources (and between the external sources themselves).

\subsection{Feynman rules}
The Feynman rules for the tree-level propagators are obtained by inverting the quadratic part of the action \eqref{action}. For the massless case the non-vanishing ones are given by \cite{Capri:2018ir} (the `q' label above the fermionic line stands for quark).
\begin{equation} \label{prop}
\begin{split}
\langle \bar \psi^i(p) \psi^j(-p)  \rangle_0 = 
%&i\feynmandiagram [ horizontal=a to b] { a  -- [fermion, momentum = \(\scriptstyle{ p}\)] b  }; j
&\,\,{\scriptstyle{i}}\, %\feynmandiagram [ horizontal=a to b] { a  -- [fermion, edge label = \(\scriptstyle{ q}\)] b  };
\begin{tikzpicture}
%\vertex (a) at (0,0); \vertex (b) at (2,0);
%\propag[chagho] (a) to [edge label = $\scriptstyle{\eta}$] (b);
\draw[->-=.5, thick, line width=0.16mm] (0,0) -- node[above] {$\scriptstyle{q}$}  (2,0);
\end{tikzpicture}\, {\scriptstyle{j}}\,\,
 = \frac{-i\slashed{p}}{p^2} \delta^{ij}, \\
\langle A^a_\mu (p) A^b_\nu(-p) \rangle_0 =
&\,\,{\scriptstyle{^a_\mu}}\, %\feynmandiagram [horizontal=a to b] { a  -- [gluon, edge label = \(\scriptstyle{ A}\)] b  };
\begin{tikzpicture}
\begin{feynman}
\vertex (a) at (0,0); \vertex (b) at (2,0);
\diagram* {
(a) -- [gluon, edge label = $\scriptstyle{ A}$]  (b) ,
};
%\propag[gluon] (a) to [edge label = $\scriptstyle{A}$] (b);
\end{feynman}
\end{tikzpicture}
\, {\scriptstyle{^b_\nu}}\,\,
= \frac{1}{p^2} \left(\delta_{\mu \nu} +(\alpha-1)\frac{p_\mu p_\nu}{p^2} \right) \delta^{ab}, \\
\langle A^a_\mu (p) b^b(-p) \rangle_0 =
&\,\,{\scriptstyle{^a_\mu}}\, \feynmandiagram [layered layout, horizontal=a to b] { a  -- [gluon, edge label = \(\scriptstyle{ A}\)] b  
-- [dash pattern=on 3pt off 2pt on \the\pgflinewidth off 2pt, edge label =  \(\scriptstyle{ b}\)] c};\, {\scriptstyle{^b}}\,\,
= \frac{p_\mu}{p^2}\,  \delta^{ab},\\
\langle A^a_\mu (p) \xi^b(-p) \rangle_0 =
&\,\,{\scriptstyle{^a_\mu}}\, \feynmandiagram [layered layout, horizontal=a to b] { a  -- [gluon, edge label = \(\scriptstyle{ A}\)] b  
-- [scalar, edge label =  \(\scriptstyle{ \xi}\)] c};\, {\scriptstyle{^b}}\,\,
 =-i\alpha \frac{p_\mu}{p^4}\, \delta^{ab}, \\
\langle b^a (p) \xi^b(-p) \rangle_0 = 
&\,\,{\scriptstyle{a}}\, \feynmandiagram [layered layout, horizontal=a to b] { a  -- [dash pattern=on 3pt off 2pt on \the\pgflinewidth off 2pt, edge label = \(\scriptstyle{ b}\)] b  
-- [scalar, edge label =  \(\scriptstyle{ \xi}\)] c};\, {\scriptstyle{b}}\,\,
=\frac{i}{p^2} \, \delta^{ab}, \\
\langle \xi^a(p) \xi^b(-p) \rangle_0 =
&\,\,{\scriptstyle{a}}\, %\feynmandiagram [ horizontal=a to b] { a  -- [scalar, edge label = \(\scriptstyle{ \xi}\)] b  };
\begin{tikzpicture}
\draw[dash pattern=on 3pt off 3pt] (0,0) -- node[above] {$\scriptstyle{\xi}$}  (2,0);
\end{tikzpicture}
\, {\scriptstyle{b}}\,\,
= \frac{\alpha}{p^4} \, \delta^{ab}, \\
\langle \xi^a(p) \tau^b(-p) \rangle_0 =
&\,\,{\scriptstyle{a}}\, \feynmandiagram [layered layout, horizontal=a to b] { a  -- [scalar, edge label = \(\scriptstyle{ \xi}\)] b  
-- [boson, edge label =  \(\scriptstyle{ \tau}\)] c};\, {\scriptstyle{b}}\,\,
= \frac{1}{p^2}\, \delta^{ab}\\
\langle \bar c^a(p) c^b(-p) \rangle_0 = 
&\,\,{\scriptstyle{a}}\, %\feynmandiagram [ horizontal=a to b] { a  -- [ghost, edge label = \(\scriptstyle{ c}\)] b  };
\begin{tikzpicture}
%\vertex (a) at (0,0); \vertex (b) at (2,0);
%\propag[chagho] (a) to [edge label = $\scriptstyle{\eta}$] (b);
\draw[->-=.5, dash pattern=on \pgflinewidth off 2pt, thick] (0,0) -- node[above] {$\scriptstyle{c}$}  (2,0);
\end{tikzpicture}
\, {\scriptstyle{b}}\,\,
= \frac{1}{p^2}\, \delta^{ab}\\
\langle \bar \eta^a(p) \eta^b(-p) \rangle_0 =
&\,\,{\scriptstyle{a}}\, 
\begin{tikzpicture}
%\vertex (a) at (0,0); \vertex (b) at (2,0);
%\propag[chagho] (a) to [edge label = $\scriptstyle{\eta}$] (b);
\draw[->-=.5, dash pattern=on \pgflinewidth off 2pt, thick] (0,0) -- node[above] {$\scriptstyle{\eta}$}  (2,0);
\end{tikzpicture}
\, {\scriptstyle{b}}\,\,
=\frac{1}{p^2}\, \delta^{ab},
\end{split}
\end{equation}
where the mixing fields propagators result from the fact that the quadratic part in the action is not diagonal. The auxiliary field $b^a$, despite appearing in mixing propagators, does not show up in loop diagrams, since its own propagator vanishes in any gauge and it does not interact with any other field.

The Feynman rules for the vertices that will be relevant for the present calculation, which add to the usual vertices of QCD and arise from the action $S_h$ in \eqref{action} and from $S_J$ in \eqref{source}, have been obtained using  {\sf Mathematica} package {\sf FeynRules} \cite{Alloul:2014}, and are given by (a total momentum conservation is implicitly understood)
\begin{equation}
\begin{split}
\langle A^a_\mu(p) \tau^b(q) \xi^c(k) \rangle_0 & = %-ig f^{abc} q_\mu, \\
\raisebox{-2.1em}{
\begin{tikzpicture}
\begin{feynman}
\vertex (a) at (0,0) {\!\!\!\!\(\scriptstyle{^a_\mu}\)} ; \vertex (b) at (1.5,0);
\vertex (c) at (2.5, 0.6) {\!\!\!\!$\scriptstyle{^c}$}; \vertex (d) at (2.5, -0.6) {\!\!$\scriptstyle{^b}$};
\diagram* {
(a)  -- [gluon, edge label = $\scriptscriptstyle{A}$, reversed momentum = {[arrow shorten=0.22mm,arrow distance=2mm] $\scriptstyle{p}$}]  (b) ,
(b) -- [scalar, edge label = $\scriptscriptstyle{\xi}$, momentum = {[arrow shorten=0.23mm,arrow distance=1.2mm] $\scriptstyle{k}$}]  (c) ,
(b) -- [photon, edge label = $ \scriptscriptstyle{\tau}$, near end, momentum = {[arrow shorten=0.23mm,arrow distance=1.2mm] $\scriptstyle{q}$}]  (d) ,
};
\end{feynman}
\end{tikzpicture}} = -ig f^{abc} q_\mu, \\
\langle \tau^a(p) \xi^b(q) \xi^c(k) \rangle_0 & = 
\raisebox{-2.1em}{
\begin{tikzpicture}
\begin{feynman}
\vertex (a) at (0,0) {\!\!\!\!\(\scriptstyle{^a}\)} ; \vertex (b) at (1.5,0);
\vertex (c) at (2.5, 0.6) {\!\!\!\!$\scriptstyle{^c}$}; \vertex (d) at (2.5, -0.6) {\!\!$\scriptstyle{^b}$};
\diagram* {
(a)  -- [photon, edge label = $\scriptscriptstyle{\tau}$, reversed momentum = {[arrow shorten=0.22mm,arrow distance=2mm] $\scriptstyle{p}$}]  (b) ,
(b) -- [scalar, edge label = $\scriptscriptstyle{\xi}$, momentum = {[arrow shorten=0.23mm,arrow distance=1.2mm] $\scriptstyle{k}$}]  (c) ,
(b) -- [scalar, edge label = $ \scriptscriptstyle{\xi}$, near end, momentum = {[arrow shorten=0.23mm,arrow distance=1.2mm] $\scriptstyle{q}$}]  (d) ,
};
\end{feynman}
\end{tikzpicture}} = 
\frac{g}{2} f^{abc}\, p{ \bf{\cdot}} (k-q), \\
\langle \bar \eta^a(p) \eta^b(q) A^c_\mu(k) \rangle_0 & = 
\raisebox{-2.1em}{
\begin{tikzpicture}
\begin{feynman}
\vertex (a) at (0,0) {\!\!\!\!\(\scriptstyle{^c_\mu}\)} ; \vertex (b) at (1.5,0);
\vertex (c) at (2.5, 0.6) {\!\!\!\!$\scriptstyle{^a}$}; \vertex (d) at (2.5, -0.6) {\!\!$\scriptstyle{^b}$};
\diagram* {
(a)  -- [gluon, edge label = $\scriptscriptstyle{A}$, reversed momentum = {[arrow shorten=0.22mm,arrow distance=2mm] $\scriptstyle{k}$}]  (b) ,
(b) -- [ghost,->-=.5, edge label = $\scriptscriptstyle{\bar \eta}$, momentum = {[arrow shorten=0.23mm,arrow distance=1.2mm] $\scriptstyle{p}$}]  (c) ,
(b) -- [ghost,-<-=.5, edge label = $ \scriptscriptstyle{\eta}$, near end, momentum = {[arrow shorten=0.23mm,arrow distance=1.2mm] $\scriptstyle{q}$}]  (d) ,
};
\end{feynman}
\end{tikzpicture}} = 
i g f^{abc} p_\mu, \\
\langle \bar \eta^a(p) \eta^b(q) \xi^c(k) \rangle_0 & = 
\raisebox{-2.1em}{
\begin{tikzpicture}
\begin{feynman}
\vertex (a) at (0,0) {\!\!\!\!\(\scriptstyle{^c}\)} ; \vertex (b) at (1.5,0);
\vertex (c) at (2.5, 0.6) {\!\!\!\!$\scriptstyle{^a}$}; \vertex (d) at (2.5, -0.6) {\!\!$\scriptstyle{^b}$};
\diagram* {
(a)  -- [scalar, edge label = $\scriptscriptstyle{\xi}$, reversed momentum = {[arrow shorten=0.22mm,arrow distance=2mm] $\scriptstyle{k}$}]  (b) ,
(b) -- [ghost,->-=.5, edge label = $\scriptscriptstyle{\bar \eta}$, momentum = {[arrow shorten=0.23mm,arrow distance=1.2mm] $\scriptstyle{p}$}]  (c) ,
(b) -- [ghost,-<-=.5, edge label = $ \scriptscriptstyle{\eta}$, near end, momentum = {[arrow shorten=0.23mm,arrow distance=1.2mm] $\scriptstyle{q}$}]  (d) ,
};
\end{feynman}
\end{tikzpicture}} = 
-g f^{abc} p{ \bf{\cdot}} k, \\
\langle \bar \psi^i(p) J_q^{1 j}(q) \xi^a(k) \rangle_0 & = 
\raisebox{-2.1em}{
\begin{tikzpicture}
\begin{feynman}
\vertex (a) at (0,0) {\!\!\!\!\(\scriptstyle{^a}\)} ; \vertex (b) at (1.5,0);
\vertex (c) at (2.5, 0.6) {\!\!\!\!$\scriptstyle{^i}$}; \vertex (d) at (2.5, -0.6) {\!\!$\scriptstyle{^j}$};
\diagram* {
(a)  -- [scalar, edge label = $\scriptscriptstyle{\xi}$, reversed momentum = {[arrow shorten=0.22mm,arrow distance=2mm] $\scriptstyle{k}$}]  (b) ,
(b) -- [plain, ->-=.5, edge label = $\scriptscriptstyle{\bar q}$, momentum = {[arrow shorten=0.23mm,arrow distance=1.2mm] $\scriptstyle{p}$}]  (c) ,
(b) -- [plain,-<-=.5, edge label = $ \scriptscriptstyle{J^1_q}$, near end, pos=0.8, momentum = {[arrow shorten=0.23mm,arrow distance=1.2mm] $\scriptstyle{q}$}]  (d) ,
};
\end{feynman}
\end{tikzpicture}} = 
-ig T^a_{ij}, \\
\langle \bar \psi^i(p) J_q^{2 j}(q) \xi^{a}(k) \xi^b(r)\rangle_0 & = 
\raisebox{-2.1em}{
\begin{tikzpicture}
\begin{feynman}
\vertex (a) at (0.5,0.6) {\!\!\!\!\(\scriptstyle{^a}\)} ; \vertex (b) at (1.5,0);
\vertex (c) at (0.5, -0.6) {\!\!\!\!$\scriptstyle{^b}$}; \vertex (d) at (2.5, 0.6) {\!\!$\scriptstyle{^i}$};
\vertex (e) at (2.5, -0.6) {\!\!$\scriptstyle{^j}$};
\diagram* {
(a)  -- [scalar, edge label = $\scriptscriptstyle{\xi}$, reversed momentum = {[arrow shorten=0.22mm,arrow distance=1.2mm] $\scriptstyle{k}$}]  (b) ,
(c)  -- [scalar, edge label = $\scriptscriptstyle{\xi}$, near start, reversed momentum = {[arrow shorten=0.22mm,arrow distance=1.2mm] $\scriptstyle{r}$}]  (b) ,
(b) -- [plain, ->-=.5, edge label = $\scriptscriptstyle{\bar q}$, momentum = {[arrow shorten=0.23mm,arrow distance=1.2mm] $\scriptstyle{p}$}]  (d) ,
(b) -- [plain,-<-=.5, edge label  = $ \scriptscriptstyle{J^2_q}$, near end, pos=0.8, momentum = {[arrow shorten=0.23mm,arrow distance=1.2mm] $\scriptstyle{q}$}]  (e) ,
};
\end{feynman}
\end{tikzpicture}} = 
\frac{g^2}{2} \left(T^a T^b + T^b T^a \right)_{ij}.
%\langle A^a_\mu(p) \tau^b (q) \xi^c(k) \xi^d (l) \rangle_0 & = i \frac{g^2}{2} q_\mu \left(f^{ade}f^{ebc} + f^{ace}f^{ebd} \right), \\
%\langle \tau^a(p) \xi^b(q) \xi^c(k) \xi^d(l) \rangle_0 & = \frac{g^2}{6} \left[f^{ade}f^{ebc} p{\bf \cdot}(k-q) + f^{ace}f^{ebd}p{\bf \cdot}(l-q)+f^{abe}f^{ecd}p {\bf \cdot}(l-k) \right].
\end{split}
\end{equation}
The aforementioned sources can also be introduced into {\sf FeynRules} to introduce the external lines of the relevant composite operators.

\subsection{One loop quark LKFT} \label{lkftone}
At the one loop level the only contributions to the quark LKFT come from $2 \langle \bar \psi(p) \Delta_1(-p) \rangle$ (or equivalently $2 \langle \bar \Delta_1(p) \psi(-p) \rangle$) and $\langle \bar \Delta_1(p) \Delta_1(-p) \rangle$. They are given by the following diagrams that, in the massless case, can be easily evaluated in arbitrary $D$ dimensions

\begin{align} \label{lkf1l}
&\includegraphics[ width=0.35 \paperwidth]{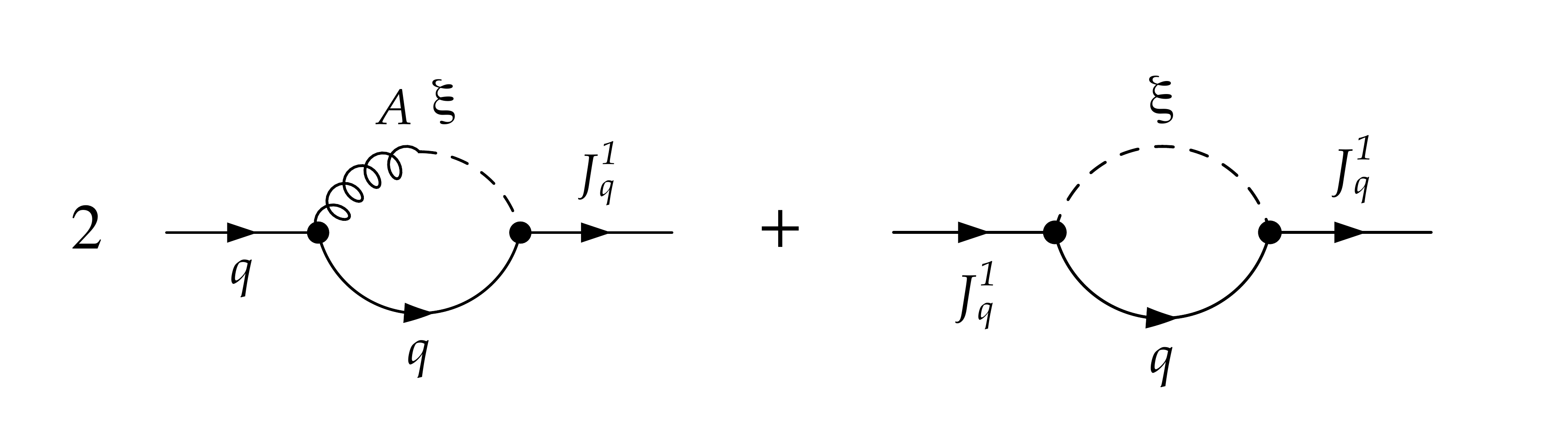} \\
&= \frac{g^2 \alpha  C_F}{(4 \pi)^{\frac{D}{2}}}\frac{i \slashed{p}}{(p^2)^{3-\frac{D}{2}}}   \left[\frac{2 \Gamma^2 \left(\frac{D}{2} \right) \Gamma\left(1-\frac{D}{2}  \right)}{\Gamma\left(D-2 \right)} - \frac{ \Gamma^2 \left(\frac{D}{2}\right) \Gamma\left(1-\frac{D}{2} \right)}{\Gamma\left(D-2 \right)} \right] \nonumber
= \frac{g^2 \alpha  C_F}{(4 \pi)^{\frac{D}{2}}}\frac{i \slashed{p}}{(p^2)^{3-\frac{D}{2}}}  \frac{ \Gamma^2 \left(\frac{D}{2} \right) \Gamma\left(1-\frac{D}{2}  \right)}{\Gamma\left(D-2 \right)}.
\end{align}
We stress that the LKFT \eqref{lkft} involves connected correlation functions, instead of 1PI ones. In order to combine the two diagrams in \eqref{lkf1l}, it is therefore necessary to multiply the first diagram by the external quark propagator (the external sources do not propagate). This will be particularly important at the two loop level (see the following subsection), where there will be contributions from connected reducible diagrams. The result \eqref{lkf1l} coincides, apart from a minus sign that has to be taken into account from \eqref{rest}, with the one loop correction to the massless quark propagator \cite{Grozin:2005}. In principle there would also be a contribution from $\langle \bar \psi(p) \Delta_2(-p) \rangle $, but it consists of a Stueckelberg tadpole diagram, which vanishes in dimensional regularization, as mentioned earlier.

This simple one loop example already allows to appreciate the main features of the present approach: on one hand the introduction of new propagators and vertices causes a proliferation of diagrams (see \cite{DeMeerleer:2020} for another one loop example concerning the LKFT of the gluon propagator). Here we have to calculate two diagrams instead of the single diagram representing the quark self-energy. On the other hand, the diagrams that make up the LKFT are generally simpler to evaluate than the traditional ones. The diagrams in \eqref{lkf1l} have in fact a simpler structure than the one loop quark self-energy, since both the Stueckelberg and the mixed propagators have simpler expressions than the gluon propagator (see \eqref{prop}). Both these facts are particularly accentuated at the two loop level. In practice, it takes only about half the time to compute the two loop gauge dependent part using LKFT compared to the traditional (direct) computation of the full two loop quark propagator in generic covariant gauge. Although this may sound reasonable, assuming that the gauge dependent part represents about half of the information contained in the full propagator, it is rather remarkable considering the great number of diagrams contained in the LKFT. In general, this suggests that in situations where one is interested in the gauge dependent part of an $n$-point function alone, whether because the gauge independent part is already known, or with the intent of checking the eventual cancellation of the gauge dependent parts of the correlation functions that build up a physical observable, the LKFT computational strategy offers a valuable (faster) alternative to the standard methodology.

\subsection{Two loop quark LKFT}
At the two loop level, all the correlation functions in \eqref{rest} contribute to the gauge dependent part of the quark propagator. The diagrams involved have been generated using  {\sf Mathematica} package {\sf FeynArts} \cite{Hahn:2000} and they have been calculated using {\sf FeynCalc} \cite{Shtabovenko:2016}. In particular, the tensor reduction into scalar integrals (see Appendix \ref{scalar}) has been realized using {\sf Tarcer} \cite{Mertig:1998}.

\begin{figure}
%\hspace*{-5cm}
\begin{subfigure}{\textwidth}
%\centering
\includegraphics[page=1, width=0.7 \paperwidth]{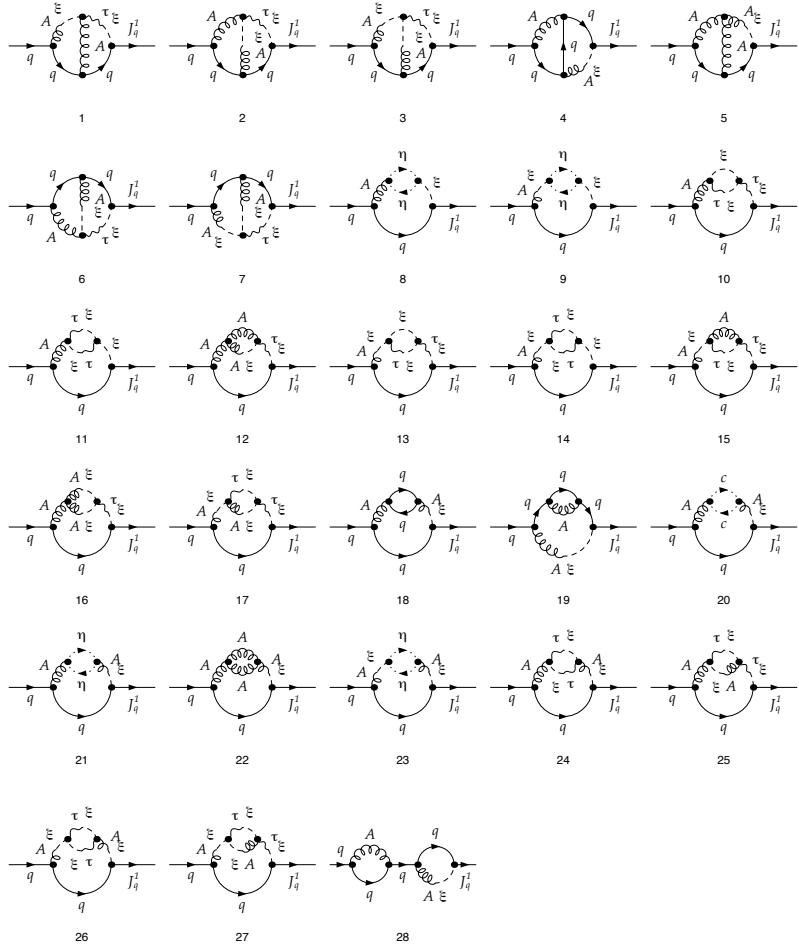}
\end{subfigure}
%\hspace{0.2cm}
%\makebox[0pt] {
\begin{subfigure}{\textwidth}
\vspace{-8.7cm}
\includegraphics[page=2, width=0.7 \paperwidth]{diagqJ12L}
\end{subfigure}
%}%\vspace{-1cm}
%\newline
\vspace{-13cm}
\caption{Feynman diagrams contributing to $\langle \bar \Delta_1(p) \psi(-p) \rangle$ at two loop level. }
    \label{diagqJ12L}
\end{figure}

The diagrams contributing to $\langle \bar \Delta_1(p) \psi(-p) \rangle^{(2)}$ are shown in FIG.~\ref{diagqJ12L} (the vanishing tadpole diagrams have been omitted). A few remarks are in order: as mentioned earlier, despite the large number of them, each diagram has a simpler structure than the ordinary diagrams present in the two loop expansion of the quark self-energy, because of the internal scalar propagators and the internal mixed $A_\mu \text{-}\xi$ propagator that acts like a longitudinal projector. In particular, it is noteworthy that the diagram 18, with the nested quark loop, vanishes automatically, since the transverse fermion loop is attached to the longitudinal propagator $A_\mu \text{-}\xi$. This is desirable, since the vacuum polarization diagrams in ordinary perturbation theory do not contribute to the gauge dependent part of the propagator. It is also remarkable that the four diagrams (8, 9, 21, 23) containing loops of the new ghost fields $\eta,\, \bar \eta$ cancel among each others. These loops were essential in proving that the longitudinal LKFT for the gluon propagator vanished at the one loop level \cite{DeMeerleer:2020}. Finally, one has to include the last diagram, which is one particle reducible, since the LKFT expansion has been derived for the connected reducible correlation functions.

\begin{figure}
%\centering
\includegraphics[page=1, width=0.4 \paperwidth]{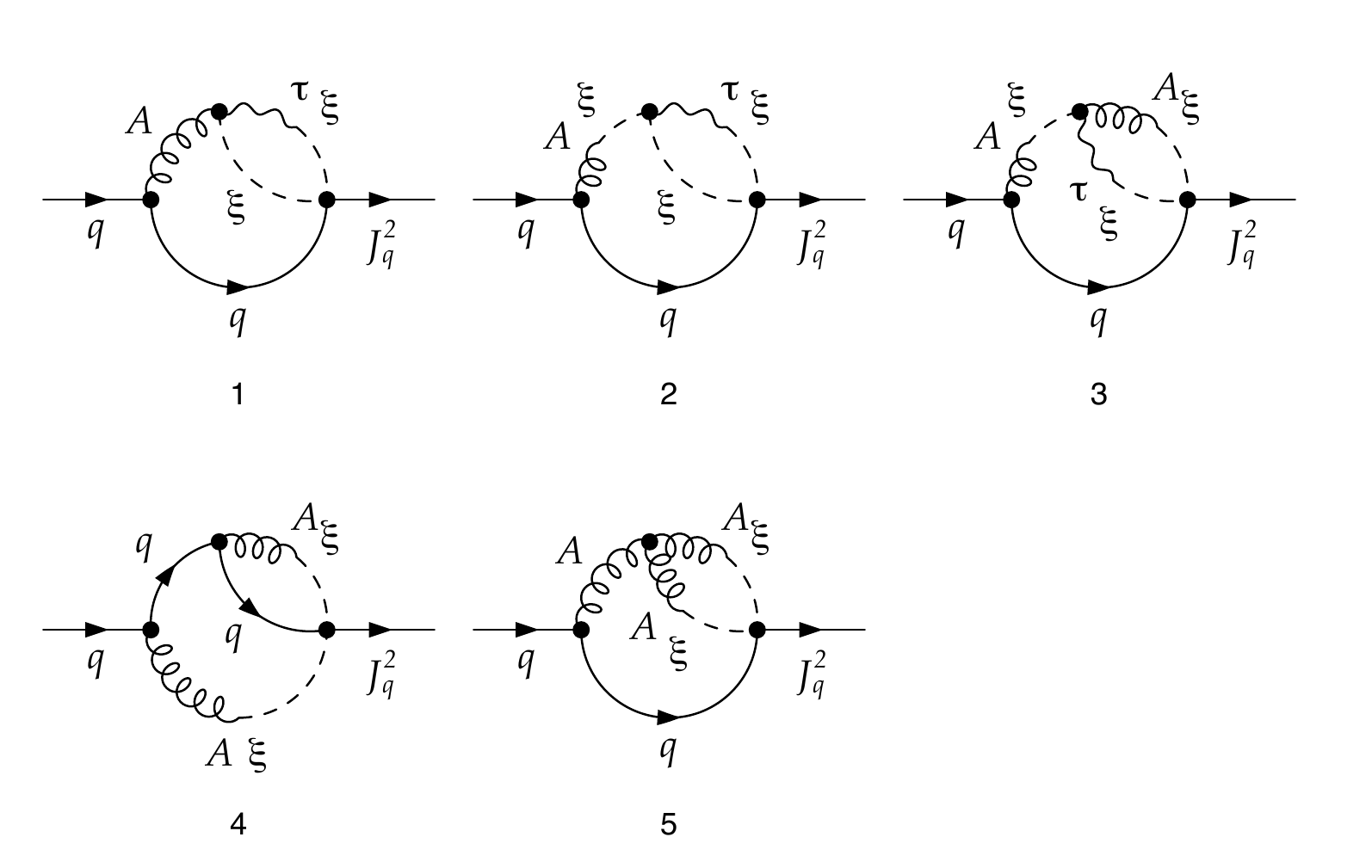}
\caption{Feynman diagrams contributing to $\langle \bar \Delta_2(p) \psi(-p) \rangle$ at two loop level. }
    \label{diagqJ22L}
\end{figure}

Adding up the diagrams we obtain (we omit the conserved color indices)
\begin{equation}
\begin{split}
\langle \bar \Delta_1(p) \psi(-p) \rangle^{(2)} & = \frac{i\slashed{p} }{(p^2)^{5-D}}\frac{g^4 \alpha C_F}{(4\pi)^D} (D-1) \Gamma^3 \left(\frac{D}{2} \right) \Gamma \left(1-D \right) \Gamma \left(1-\frac{D}{2} \right) \left \{  \frac{4\alpha C_F(D-3)}{ \Gamma \left(3-\frac{D}{2} \right)  \Gamma \left(\frac{3D}{2}-4 \right)} \right. \\
& \hspace{-2.5cm}\left.+ \frac{C_A}{D-4} \left[\frac{\alpha(D-4)((D-9)D+7)-D(D-6)(D-1)-16}{2\Gamma \left(4-\frac{D}{2} \right) \Gamma \left(\frac{3D}{2} -4 \right)}  -\frac{2 \Gamma \left(\frac{D}{2} \right) \Gamma \left(1-\frac{D}{2} \right)}{\Gamma \left(D \right) \Gamma \left(1-D \right) \Gamma \left(D-3 \right)} \right] \right\}.
\end{split}
\end{equation}
The diagrams contributing to $\langle \bar \Delta_2(p) \psi(-p) \rangle^{(2)}$ are shown in FIG.~\ref{diagqJ22L}. Of those five diagrams with the same topology, only the diagram 4 with a second quark-gluon vertex is non-vanishing. The other ones are zero due to the combination of the two vertices connected by two internal propagators, the one attached to the external source being symmetric in the color indices and the other one being antisymmetric. The expression of diagram 4 is given by
\begin{equation}
\langle \bar \Delta_2(p) \psi(-p) \rangle^{(2)} = \frac{i\slashed{p}g^4\alpha^2 C_F(C_A-4C_F)}{(p^2)^{5-D} (4\pi)^D}\frac{\Gamma\left(\frac{D}{2} \right)\Gamma^2\left(\frac{D}{2}-2 \right) \Gamma(5-D)}{8\Gamma\left(\frac{3D}{2}-4 \right)}.
\end{equation}

\begin{figure}
%\hspace*{-5cm}
\vspace{-3.7cm}
\begin{subfigure}{\textwidth}
%\centering
\includegraphics[page=1, width=0.7 \paperwidth]{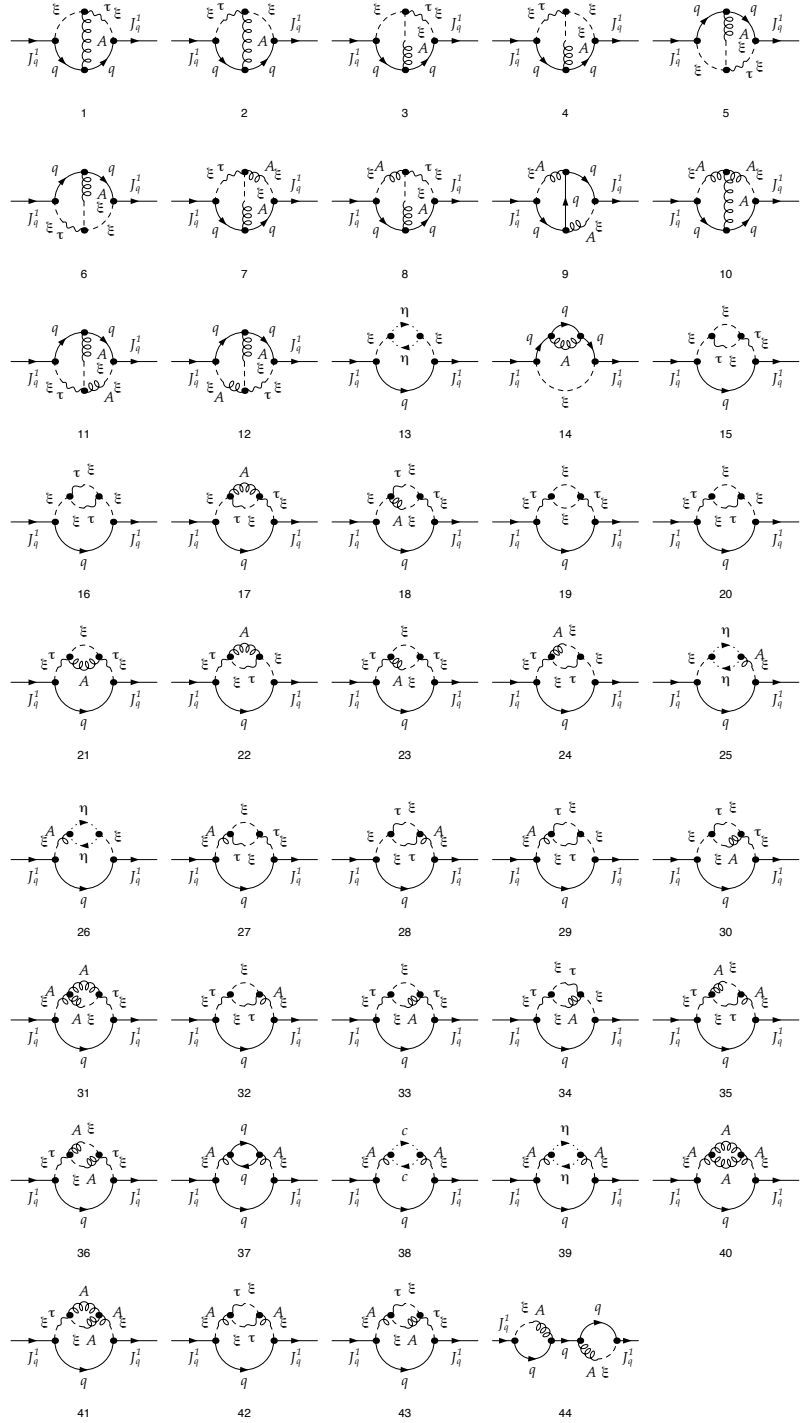}
\end{subfigure}
%\hspace{0.2cm}
%\makebox[0pt] {
\begin{subfigure}{\textwidth}
\vspace{-8.7cm}
\includegraphics[page=2, width=0.7 \paperwidth]{diagJ1J12L}
\end{subfigure}
%}%\vspace{-1cm}
%\newline
\vspace{-6.7cm}
\caption{Feynman diagrams contributing to $\langle \bar \Delta_1(p) \Delta_1(-p) \rangle$ at two loop level. }
    \label{diagJ1J12L}
\end{figure}

The diagrams contributing to $\langle \bar \Delta_1(p) \Delta_1(-p) \rangle^{(2)}$ are shown in FIG.~\ref{diagJ1J12L}. Again, diagram 37, containing a quark loop, vanishes automatically; and as before, the diagrams containing the $\eta, \, \bar \eta$ loops (13, 25, 26, 39) cancel among each other. Also in this set of diagrams the last one is reducible. These 44 diagrams sum up to give the following expression
\begin{equation}
\begin{split}
\langle \bar \Delta_1(p) \Delta_1(-p) \rangle^{(2)} & = \frac{i\slashed{p} }{(p^2)^{5-D}} \frac{g^4 \alpha C_F}{(4\pi)^D} \Gamma^3\left( \frac{D}{2}\right) \Gamma\left(1-\frac{D}{2} \right) \left \{\frac{C_A  \Gamma\left(1-\frac{D}{2} \right)}{16 \Gamma^2(D-2)} \left[ \frac{\Gamma(D)\Gamma(1-D) \Gamma(D-2) }{ \Gamma^2\left(4-\frac{D}{2} \right)  \Gamma\left(\frac{3D}{2}-4 \right)} \right. \right. \\
& \times (D-6)\left(\alpha(D-4)(2(D-9)D+35)-(D-6)(D-1)D-16 \right) \\
&\left. \left. + \frac{32 \Gamma\left(\frac{D}{2} \right)  (D-3)}{(D-4)(D-2)} \right] -  \frac{8 \alpha C_F \Gamma(1-D)(D-1)(D-3)}{\Gamma\left(3-\frac{D}{2} \right)\Gamma\left(\frac{3D}{2}-4 \right) } \right \}.
\end{split}
\end{equation}

The diagrams contributing to $\langle \Delta_2(p) \Delta_1(-p) \rangle^{(2)}$ have the same topology as the diagrams contributing to  $\langle \Delta_2(p) \psi(-p) \rangle^{(2)}$, and also in this case the only non-vanishing diagram is the one where the vertex attached to external source $J_q^2$ is connected to a quark-gluon vertex. The result is
\begin{equation}
\vspace{-5cm}
\langle \bar   \Delta_2(p) \Delta_1(-p) \rangle^{(2)}=\vspace{5cm}\includegraphics[ width=0.15 \paperwidth, valign=c]{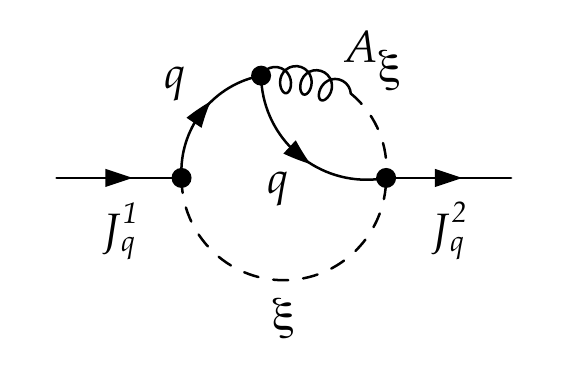} =-  \frac{i\slashed{p}g^4\alpha^2 C_F(C_A-4C_F)}{(p^2)^{5-D} (4\pi)^D} \frac{\Gamma(5-D)\Gamma^2\left(\frac{D}{2}-2 \right)\Gamma\left(\frac{D}{2} \right)}{4\Gamma\left(\frac{3D}{2}-4 \right)}.
\end{equation}
Finally, the correlation function $\langle \bar \Delta_2(p) \Delta_2(-p) \rangle^{(2)}$ is also given by one diagram
\begin{equation}
\vspace{-5cm}
\langle \bar   \Delta_2(p) \Delta_2(-p) \rangle^{(2)}=\vspace{5cm}\includegraphics[ width=0.15 \paperwidth, valign=c]{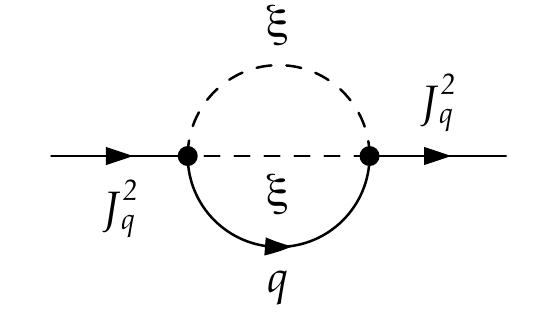} =  \frac{i\slashed{p}g^4\alpha^2 C_F(C_A-4C_F)}{(p^2)^{5-D} (4\pi)^D} \frac{\Gamma(D)\Gamma(1-D)\Gamma^2\left(\frac{D}{2}-2 \right)\Gamma\left(\frac{D}{2} \right)}{8\Gamma(D-4)\Gamma\left(\frac{3D}{2}-4 \right)}.
\end{equation}
Putting all the pieces together, we obtain the expression for the gauge dependent part of the fermionic massless propagator \eqref{rest} in momentum space at two loop level ${\cal R}_{\alpha}^{(2)}(p)$

\begin{equation}
\begin{split}
\label{result}
{\cal R}_{\alpha}^{(2)}(p) & = - 2\langle \bar \Delta_1(p)  \psi(-p) \rangle^{(2)} - 2\langle \bar \Delta_2(p) \psi(-p) \rangle^{(2)} - 2\langle \bar \Delta_2(p) \Delta_1(-p) \rangle^{(2)} -\langle \bar \Delta_1(p) \Delta_1(-p) \rangle^{(2)}-\langle \bar \Delta_2(p) \Delta_2(-p) \rangle^{(2)} \\
& =  -\frac{i\slashed{p}}{(p^2)^{5-D}} \frac{g^4\alpha C_F}{(4\pi)^D} \Gamma^2\left(\frac{D}{2} \right)  \left \{\frac{C_A}{2} \Gamma \left(\frac{D}{2} \right)\Gamma\left(1-\frac{D}{2} \right) \left[-\frac{\Gamma \left(\frac{D}{2} \right)\Gamma^2\left(1-\frac{D}{2} \right) (D-3)}{\Gamma\left(3-\frac{D}{2} \right)\Gamma^2(D-2)} \right. \right. \\
&+ \left. \frac{\Gamma(1-D) (D-1)(\alpha(D-4)((D-9)D+16)-2((D-6)(D-1)D+16))}{2(D-4)\Gamma\left(4-\frac{D}{2} \right)\Gamma\left(\frac{3D}{2}-4 \right) } \right] \\
& + \left. 2C_F \alpha \frac{\Gamma(1-D)\Gamma\left(\frac{D}{2} -2 \right)(D-1)(D-3)}{\Gamma\left(\frac{3D}{2}-4 \right)} \right\}.
\end{split}
\end{equation}

\subsection{$\overline{\text{MS}}$ renormalization constant}
In order to check the validity of the result \eqref{result}, we are going to extract from it the gauge dependent part of the quark renormalization constant $Z_2$ in the $\overline{\text{MS}}$ scheme in $D=4$, up to the two loop level, and compare it to the one reported in \cite{Fleischer:1999}.
The gauge dependent part of the bare connected propagator, up to the two loop level, is given by
\begin{equation}
\label{bareprop}
{\cal R}_{\alpha}(p) = -\frac{i \slashed{p}}{p^2} \frac{\bar g^2 \alpha}{(4 \pi)^{\frac{D}{2}}}\left( \frac{p^2}{\mu^2} \right)^{-\epsilon}\left[e^{\gamma \epsilon}f_{1}(\epsilon) + \frac{\bar g^2 e^{2 \gamma \epsilon}}{(4\pi)^\frac{D}{2}}\left(\frac{p^2}{\mu^2}\right)^{-\epsilon} f_{2}(\epsilon)  \right],
\end{equation}
where $\epsilon = \frac{4-D}{2}$ and $f_{1}(\epsilon)$ and $f_{2}(\epsilon)$ are respectively the one loop and two loop contributions that are factorized from \eqref{lkf1l} (with a minus sign) and \eqref{result}. $\bar g^2 = e^{-\gamma \epsilon} \mu^{-2\epsilon} g^2$ is the dimensionless coupling constant written in terms of the arbitrary scale $\mu$ and the Euler constant $\gamma$. The coefficients in the expansions of $f_1(\epsilon)$ and $f_2(\epsilon)$
\begin{align}
e^{\epsilon \gamma} f_1(\epsilon) &= \frac{1}{\epsilon} (c_{10}+c_{11} \epsilon + \cdots), \\
e^{2\epsilon \gamma} f_2(\epsilon) &= \frac{1}{\epsilon^2} (c_{20}+c_{21} \epsilon + \cdots) \nonumber
\end{align}
are found by expanding around $D=4-2\epsilon$  \eqref{lkf1l}, which gives $c_{10} = c_{11} = -C_F$ and \eqref{result}, implying $c_{20} = \frac{C_F}{4} \left(2\alpha C_F-(\alpha+3)C_A \right)$ and $c_{21} =\frac{C_F}{8}\left(8\alpha C_F -5(\alpha+4)C_A \right)$. Writing the coupling constant and the gauge parameter in terms of their renormalized counterparts with the corresponding renormalization constants given at one loop
\begin{align}
\bar g^2 &= \bar g^2_R Z_{g^2} = \bar g^2_R \left(1 - \beta_0 \frac{\bar g^2_R}{(4 \pi)^2} \frac{1}{\epsilon} +\cdots \right), \quad \beta_0 = -\frac{1}{3}(2n_f -11C_A), \\
\alpha &=  \alpha_R Z_3 = \alpha_R \left(1 - \frac{\gamma_{3,0}}{2} \frac{\bar g^2_R}{(4 \pi)^2} \frac{1}{\epsilon} +\cdots  \right), \quad \gamma_{3,0} = \frac{1}{3}\left(C_A(3\alpha -13) +4N_f \right) \nonumber,
\end{align}
and equating \eqref{bareprop} to $Z_2 {\cal R}_{\alpha, R}(p)$, one obtains the coefficients in the expansion of $Z_2$
\begin{align}
& Z_2 = 1+ \frac{g_R^2}{\epsilon (4 \pi)^2 } z_1 + \left(\frac{g_R^2}{\epsilon (4 \pi)^2 }\right)^2(z_{20} + z_{21}\epsilon +\cdots) +\cdots  \\ \nonumber
\Longrightarrow \, \, & z_1 = \alpha c_{10} = -\alpha C_F, \\ \nonumber
& z_{20} = \alpha \left( c_{20} - c_{10} \left(\frac{\gamma_{3,0}}{2} + \beta_0 \right) \right) = \frac{\alpha C_F}{4}\left((\alpha+3)C_A + 2\alpha C_F \right), \\ \nonumber
& z_{21} = \alpha \left(c_{21}-c_{11}\left(\frac{\gamma_{3,0}}{2} + \beta_0 + c_{10} \alpha \right) \right) = -\frac{C_F C_A}{8} \alpha(\alpha+8), \nonumber
\end{align}
which coincide with the $\alpha$-dependent parts of the coefficients reported in \cite{Fleischer:1999}.

\subsection{Resummation of the LKFT in exponential form}
In previous work \cite{DeMeerleer:2020}, we already established the underlying connection between the Nielsen identities and the LKFTs via the underlying Slavnov-Taylor identity. That both are related is no surprise, given that both govern nothing more (or less) than the precise gauge parameter dependence of generic $n$-point functions.

Specializing to the case of a connected propagator $D(p)$, the \emph{integrated} Nielsen identities imply that \cite{DeMeerleer:2020}
\begin{equation}\label{nn1}
  D_\alpha(p)= e^{N(\alpha,p)} D_{\alpha=0}(p)
\end{equation}
where the gauge dependence is automatically resummed into exponential form. We stress that, as opposed to QED, it is not clear yet how to infer this exponential resummation solely within the context of the LKFT, due to the non-trivial interactions of the Stueckelberg field in the non-Abelian case. 

In the massless case, we can write for the quark propagator
\begin{equation}\label{nn2}
\langle \bar \psi^i(p) \psi^j(-p)  \rangle  = \frac{-i\slashed{p}Z(p)}{p^2} \delta^{ij}
\end{equation}
and thence
\begin{equation}\label{nn3}
  Z_\alpha(p)= e^{N(\alpha,p)} Z_{\alpha=0}(p),
\end{equation}
while the LKFT has given us access to the rest function ${\cal R}_\alpha(p)$
\begin{equation}\label{nn4}
 {\cal R}_\alpha(p)= Z_\alpha(p)-Z_{\alpha=0}(p) = g^2 Z^{(1)}(\alpha,p)+g^4 Z^{(2)}(\alpha,p)+\ldots.
\end{equation}
This means that up to the considered two loop order, we can read off from our LKFT output the exponent $N(\alpha,p)$
\begin{equation}
N(\alpha, p) = g^2 Z^{(1)}(\alpha,p) + g^4 \left(Z^{(2)}(\alpha,p) -\frac{(Z^{(1)}(\alpha,p))^2}{2} \right) + \ldots.
\end{equation}
Such resummation was also attempted at in \cite{Aslam:2016}, but notice that the Stueckelberg field was considered free there, which is no longer tenable beyond the Abelian approximation from the viewpoint of renormalization, as confirmed by this work.

\begin{figure}
\centering
\begin{subfigure}{.5\textwidth}
  \centering
  \includegraphics[width=1.07\linewidth]{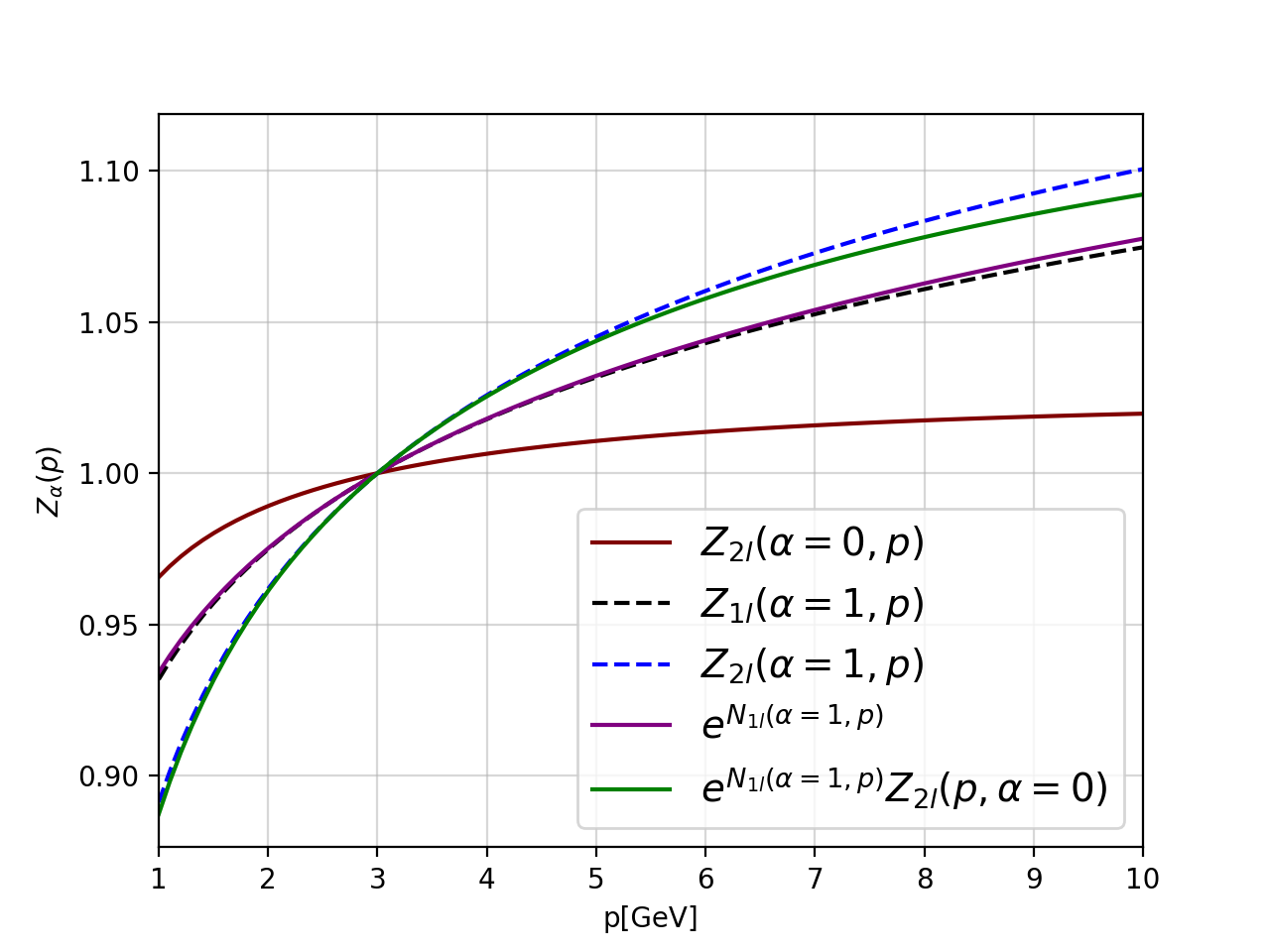}
\end{subfigure}%
\begin{subfigure}{.5\textwidth}
  \centering
  \includegraphics[width=1.07\linewidth]{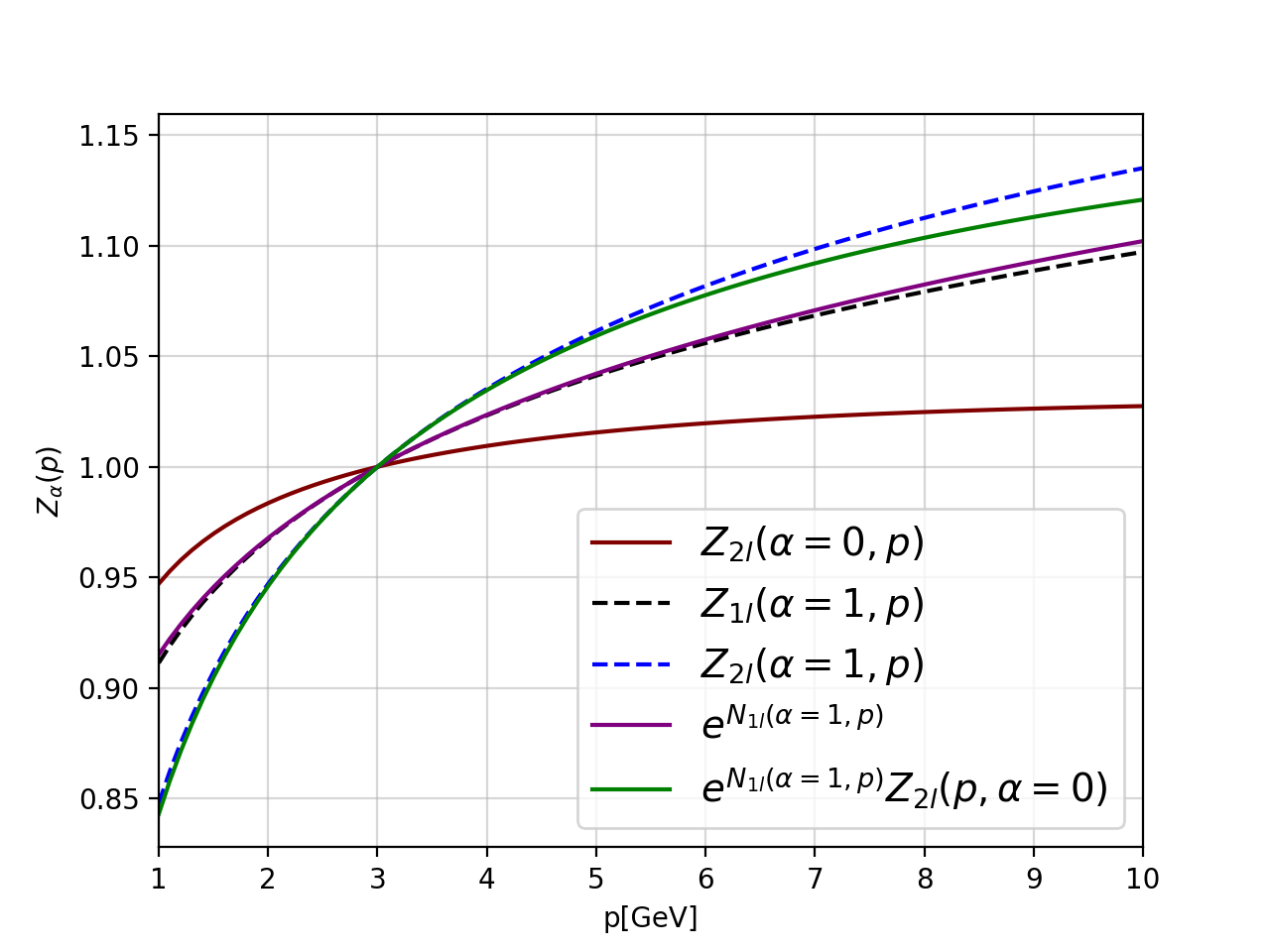}
\end{subfigure}
\caption{Quark form factor $Z_{\alpha}(p)$ at one loop ($Z_{1l}(\alpha,p)$) and two loop levels ($Z_{2l}(\alpha,p)$) in Landau ($\alpha=0$) and Feynman ($\alpha=1$) gauges, and the corresponding resummed versions $e^{N(\alpha, p)}Z_{\alpha=0}(p)$ (notice that $Z_{1l}(\alpha=0, p) = 1$). The two plots correspond to $N_f=0$ (left) and $N_f=2$ (right). }
\label{uvquark}
\end{figure}

In FIG.~\ref{uvquark}, we display the relative change w.r.t.~the Landau gauge quark form factor, both using the direct LKFT results and its resummed version. We show in particular the plots corresponding to the Feynman gauge $\alpha=1$, at both one loop and two loop levels, and the form factor in Landau gauge at two loop level (the one loop level result in Landau gauge is trivially equal to $1$). We also show the results both in the quenched case, when there are no fermions inside the loops of the quantum corrections evaluated in Landau gauge ($N_f=0$) and the case when $N_f=2$. The dependence on $N_f$ appears only at the two loop level and there are no qualitative differences between the two sets of results. We also notice that while there is a considerable shift from Landau gauge to Feynman gauge and also from one loop to two loop orders, the discrepancy between the direct results and the corresponding resummed versions is minimal, and only noticeable at high energies.

We have used a momentum subtraction (MOM) scheme here, at scale $\mu=3~\text{GeV}$ ($Z(p, \mu=3) = 1$). For the MOM coupling, by assuming a MOM scheme for the gluon and ghost propagator as well, we used its one loop expression $g^2(\mu)=\frac{(4 \pi)^2}{\beta_0 \ln \frac{\mu^2}{\Lambda^2}}$ where the $\Lambda$-scale for $N_f=0$ is borrowed from \cite{Boucaud:2009} ($\Lambda_{N_f=0}=0.42~\text{GeV}$) and for $N_f=2$ from \cite{Blossier:2010} ($\Lambda_{N_f=2}=0.54~\text{GeV}$), so that $g^2_{N_f=0}= 3.7$, $g^2_{N_f=2}=4.8$.

Since we are using perturbation theory, at sufficiently low scale a Landau pole will emerge in perturbation theory, similarly as in the $\overline{\text{MS}}$ case. As mentioned before, this can be dealt with non-perturbatively by including a gluon mass scale and appropriate renormalization scheme, cfr.~\cite{Pelaez:2021tpq,Barrios:2021cks}, see also \cite{DallOlio:2020xpu}. We hope to come back to this issue in future work in relation to the LKFTs.

\subsection{Conclusion}
In this work we corroborated the validity of a recent formalism that allows to derive the LKFTs for arbitrary correlation functions in a non-Abelian gauge theory, by explicitly deriving and calculating the LKFT for the massless quark propagator up to the two loop level. The formalism allows to separately evaluate the gauge dependent part of the propagator by expanding a gauge invariant composite operator $\psi^h$ into the original fermionic field $\psi$ and powers of a scalar Stueckelberg field. The presence of the Stueckelberg field combined with another auxiliary field that imposes the transversality of the corresponding gauge invariant composite operator $A^h_\mu$, causes a proliferation of the Feynman diagrams to be calculated which questions the practicality of the whole formalism. However, since the diagrams involved have a much simpler tensorial structure compared to the ones generated in the usual full perturbative expansion, we could verify that the evaluation of these diagrams, involved in the gauge dependent part of the propagator, is actually computationally more economical. This makes the present approach a viable strategy of computation, especially in the cases where only the gauge dependent part of a correlation function is under scrutiny.

We verified the correctness of the gauge dependent part of the quark propagator derived using the LKFT in arbitrary spacetime dimension, by deriving the gauge dependent part of the quark renormalization constant in the $\overline{\text{MS}}$, in $D=4-2\epsilon$ dimensions, and comparing to the one in the literature. We also separately calculated the gauge independent part of the quark propagator in order to compare the behavior of the full propagator in different gauges, especially in Landau ($\alpha=0$) and Feynman ($\alpha=1$) gauge and also in their resummed versions, as dictated by the Nielsen identity, where the gauge dependent content of the propagator is encoded in an exponential factor.

These promising results suggest that the formalism should be pushed forward in order to explore the massive case, where a mass term for the quark and/or the gluon is taken into account. This could shed light on the gauge dependency of these particles mass functions.

\section*{Acknowledgments}
P.~Dall'Olio was supported by a DGAPA-UNAM grant. This study was financed in part by the Coordena\c{c}\~{a}o de Aperfei\c{c}oamento de Pessoal de N\'{i}vel Superior - Brasil (Capes) - Finance Code 001. The Conselho Nacional de Desenvolvimento Cient\'{\i}fico e Tecnol\'{o}gico (CNPq-Brazil), the Funda\c{c}\~{a}o de Amparo \`{a} Pesquisa do Estado do Rio de Janeiro (FAPERJ) and the SR2-UERJ are gratefully acknowledged for financial support. S. P. Sorella is a level PQ-1 researcher under the program Produtividade em Pesquisa-CNPq, 301030/2019-7.
This research was also partly supported by Coordinaci\'on de la Investigaci\'on Cient\'ifica
(CIC) of the University of Michoacan and CONACyT,
Mexico, through Grant nos. 4.10 and CB2014-22117, respectively. T.~De Meerleer was supported by a KU Leuven FLOF grant.

\appendix
\section{Two loop scalar integrals}
\label{scalar}
The Mathematica package {\sf Tarcer} performs the tensorial reduction of two loop self-energy diagrams in terms of scalar integrals \cite{Mertig:1998}. The massless scalar integrals that are relevant for the two loop diagrams discussed in this work, are of the following type
\begin{equation}
\label{scalar}
I\left(p^2, \{u, v, r, s, t \}, \{\nu_1, \nu_2, \nu_3, \nu_4, \nu_5 \} \right)= \int \! \frac{d^Dl}{(2\pi)^D}\! \int \! \frac{d^D k}{(2\pi)^D}\! \frac{(l^2)^u (k^2)^v (p\cdot l)^r(p\cdot k)^s (l\cdot k)^t}{(l^2)^{\nu_1} (k^2)^{\nu_2}((l-p)^2)^{\nu_3}((k-p)^2)^{\nu_4}((l-k)^2)^{\nu_5}}.
\end{equation}
Among the 34 different scalar integrals that appear after the reduction of all the diagrams shown above, the majority is easily factorized into the product of two one loop scalar integrals. Only two of them are genuinely two loop integrals, and must be decomposed into factorizable integrals using the method of integration by parts \cite{Chetyrkin:1981} or by introducing the Gegenbauer polynomials \cite{Chetyrkin:1980}. The first one is (we omit the first set of indices in \eqref{scalar} being all zeros)
\begin{equation}
\begin{split}
I(p^2, \{1,1,1,1,1\}) & =  \int \! \frac{d^Dl}{(2\pi)^D}\! \int \! \frac{d^D k}{(2\pi)^D}\frac{1}{l^2k^2(l-p)^2(k-p)^2(l-k)^2}\\
&= \frac{1}{D-4} \left[I(p^2, \{1,1,1,2,0\}) +I(p^2, \{1,2,1,1,0\}) - I(p^2, \{0,2,1,1,1\})  \right. \\
&\,\,\,\,\, \left.- I(p^2, \{1,1,0,2,1\}) \right] \\
& = \frac{2(p^2)^{D-5}}{(4 \pi)^D}(D-1) \frac{\Gamma^2\left(\frac{D}{2} \right)\Gamma\left(1-\frac{D}{2} \right) }{\Gamma\left(3-\frac{D}{2} \right)}\left[\frac{2 \Gamma\left(\frac{D}{2}-2 \right)\Gamma(1-D)}{\Gamma\left(\frac{3D}{2}-5 \right)}-\frac{\Gamma^2\left(\frac{D}{2} \right)\Gamma^2\left(1-\frac{D}{2} \right)}{\Gamma(D) \Gamma(D-3)} \right].
\end{split}
\end{equation}
The second one is
\begin{equation}
\begin{split}
I(p^2, \{1,1,1,2,1\}) & =  \int \! \frac{d^Dl}{(2\pi)^D}\! \int \! \frac{d^D k}{(2\pi)^D}\frac{1}{l^2k^2(l-p)^2(k-p)^4(l-k)^2}\\
&= \frac{1}{D-5} \left[2 I(p^2, \{1,1,1,3,0\}) +I(p^2, \{1,2,1,2,0\}) - I(p^2, \{0,2,1,2,1\})  \right. \\
&\,\,\,\,\, \left.- 2I(p^2, \{1,1,0,3,1\}) \right] \\
& = \frac{8(p^2)^{D-6}}{(4\pi)^D} \frac{\Gamma^3\left(\frac{D}{2} \right)}{(D-4)(D-2)^2}\left[\frac{\Gamma\left(\frac{D}{2} \right)\Gamma^2\left(1-\frac{D}{2} \right)}{\Gamma(D-4)\Gamma(D-2)} -\frac{12(D-5)(D-1)\Gamma(1-D)}{(D-6)\Gamma\left(\frac{3D}{2}-5 \right)} \right].
\end{split}
\end{equation}

\end{document}